\documentclass{article}
\usepackage{epsfig}
\usepackage{graphics}
\usepackage{amsmath}
\usepackage{amsfonts}
\usepackage{amssymb}
\begin{document}
\title{Oscillating dark energy: a possible solution to the problem of eternal
acceleration}
\author{C. Rubano, P. Scudellaro, E. Piedipalumbo \\
Dipartimento di Scienze Fisiche, Universit\`{a} di Napoli,\\ Complesso
Universitario di Monte S. Angelo, Via Cintia, Ed. N, \\ I-80126
Napoli, Italy \\ and Istituto Nazionale di Fisica Nucleare, Sez.
Napoli, \\ Complesso Universitario di Monte Sant'Angelo, Via Cintia,
Ed. G, \\ I-80126 Napoli, Italy \and S. Capozziello \\
Dipartimento di Fisica ``E. R. Caianiello'' - Univ. di Salerno\\ and
Istituto Nazionale di Fisica Nucleare, Sez. Napoli, \\ Complesso
Universitario di Monte Sant'Angelo, Via Cintia, Ed. G, \\ I-80126
Napoli, Italy}
\date{\today}
\maketitle

\begin{abstract}
We present a dark energy model with a double exponential potential for
a minimally coupled scalar field, which allows general exact
integration of the cosmological equations. The solution can perfectly
emulate a $\Lambda$-term model at low-medium redshift, exhibits
tracking behavior, and does not show eternal acceleration.
\end{abstract}

\section{Introduction}

Among the many problems posed by the discovery of positive
acceleration of the Universe, one of the most puzzling is that of
eternal acceleration. In the case of a true cosmological constant, as
well as for most of the models proposed in the literature, the present
time accelerating era seems doomed to last forever. This creates
problems in string theory, where asymptotic flatness of spacetime is
required \cite{1}. Several possible solutions are based on cycles in
the evolution of the Universe \cite{2}, or on the introduction of
negative potentials for the scalar field \cite{3}. A much simpler
approach is obtained with a negative cosmological constant, together
with a suitable potential \cite{4}. However, this last approach poses
the same type of problems which make unsatisfactory the simple use of
a standard positive $\Lambda $. A common feature of many of these
models is that they foresee a recollapsing universe in the context of
a spatially flat metric. But a recollapsing universe can also be
obtained simply by setting $k=+1$, an option which has not been ruled
out by the BOOMERANG and MAXIMA missions, as instead people often say.
On the contrary, the most recent data by Wilkinson Microwave
Anisotropy Probe (WMAP) have reopened the way to nonflat cosmological
models \cite{5}, \cite{6}.

In this paper, we remain in the context of a flat space $\left(
k=0\right) $ and propose a very simple and attractive model with a
minimally coupled scalar field plus dust. We shall find a general
exact solution of the equations, and show that it meets all the
requests which are usually made for a dark energy model, including
tracking behavior and switch off of the acceleration. Moreover, we
shall not get a recollapsing universe.

Let us consider the following potential (in units such that $8\pi
G=1$):
\begin{eqnarray}
V(\varphi ) &=&\left( A\exp \left( \frac{1}{2}\sqrt{\frac{3}{2}}\varphi
\right) -B\exp \left( -\frac{1}{2}\sqrt{\frac{3}{2}}\varphi \right) \right)
^{2}  \nonumber \\
&=&A^{2}\exp \left( \sqrt{\frac{3}{2}}\varphi \right) +B^{2}\exp \left( -%
\sqrt{\frac{3}{2}}\varphi \right) -2AB.  \label{eq1}
\end{eqnarray}

\noindent

This potential has already been discussed in the literature (see Refs.
\cite{7}, \cite{8}, \cite{9} for instance). In Ref. \cite{8}, which is
a paper closely related to this one, the potential is not equal but
only similar to that in Eq. (\ref{eq1}). Comments on the attractor
properties of the (generic) solutions due to such a kind of potential
in connection with initial conditions and the values of model
parameters have also been given, without trying to find any exact
solution \cite{7}. On the other hand, Ref. \cite{9} inverts our point
of view, since it postulates a special scale factor behavior, so
deducing a potential form similar to the one in Eq. (\ref{eq1}
with some relevant differences with respect to it. (The main point is
the different relation among the values of the potential parameters.)
The potential (\ref{eq1}) also turns out to be a particular case of
the wider class discussed in Ref. \cite{4}. For a discussion on the
possibility of finding a general exact solution for the kind of
potentials to which the one in Eq. ( \ref{eq1}) belongs, however, see
Ref. \cite{10}.

In any case, here\ we do not present a new technique for finding
solutions; instead, we select the particular class of potentials in
Eq. (\ref{eq1}) and a treatment of the related solutions, which allows
us to obtain the desired results and puts them in greatest evidence.

Some comments are now needed on Eq. (\ref{eq1}).

(1) The second form of the potential can also be interpreted as
containing a negative cosmological constant, fine tuned with the
parameters of the potential itself. The first form shows that this is
in fact only a matter of interpretation. Such a form as a whole, of
course, still retains the right to be used as a potential in itself.

(2) In particular, if we set $A=B$, the expansion of $V(\varphi )$
around its minimum is
\begin{equation}
V(\varphi )\approx \frac{1}{2}(3B^{2})\varphi ^{2}+\frac{1}{16}
(3B^{2})\varphi ^{4}+...,  \label{eq2}
\end{equation}
showing that it can be interpreted as a massive self-interacting
potential, which is quite reasonable in the situation we are studying.
We shall see that posing $A = 0$ only affects the asymptotic value of
$\varphi $. As a matter of fact, we consider $A\neq B$ just for sake
of generality.

(3) The real problems with potential (\ref{eq1}) are the particular
value of the exponent, \ and the link between the mass term and the
self-interaction term in Eq. (\ref{eq2}). We are not really able to
explain them in terms of fundamental theory\footnote{But, about this,
see Refs. \cite{11},\cite{12} for some dimensional considerations on
exponential potential, once one wants power law solutions for the
scale factor.}. It is however intriguing that, as we shall see, the
dynamics entailed by this potential on the minisuperspace $\{a,\varphi
\}$ admits a Noether symmetry.

After presenting the cosmological solution in Sec. II, we analyze it
in Sec.\ III. Finally, Sec. IV is dedicated to some conclusive
remarks.

\section{Solution}

\bigskip

Let us consider the Einstein equations, together with the Klein-Gordon
equation for the scalar field, in the flat case

\begin{eqnarray}
3H^{2} &=&\rho _{\varphi }+\rho _{m},  \label{eq4} \\
2\frac{\ddot{a}}{a}+\left( \frac{\dot{a}}{a}\right)
^{2}+\frac{1}{2}\dot{
\varphi}^{2}-V(\varphi ) &=&0,  \label{eq5} \\
\ddot{\varphi}+3H\dot{\varphi}+V^{\prime }(\varphi ) &=&0.  \label{eq6}
\end{eqnarray}

Let us concentrate on the last two equations and put aside the first
one, which is, as is well known, a first integral. The reason for such
a choice is that these equations can be derived by the Lagrangian
\cite{13}, \cite{14}
\begin{equation}
L=3a\dot{a}^{2}-\frac{1}{2}a^{3}\dot{\varphi}^{2}+a^{3}V(\varphi ).
\label{eq7}
\end{equation}

In close analogy with Ref. \cite{8}, we now set a transformation of
variables, which is however not the same as there
\begin{equation}
a^{3}=\frac{3(u^{2}-v^{2})}{4\omega ^{2}}\quad ,\quad \varphi =\sqrt{\frac{2%
}{3}}\log \frac{u+v}{u-v}+\varphi _{\infty },  \label{eq8}
\end{equation}
where $\omega ^{2}\equiv 3AB$ and $\varphi _{\infty }\equiv \sqrt{2/3}\log
(B/A)$. With this change, we get for $L$%
\begin{equation}
L=\frac{1}{\omega ^{2}}(\dot{u}^{2}-\dot{v}^{2})+v^{2},  \label{eq9}
\end{equation}
and we see that the variable $u$ is cyclic, so that we have got a Noether
conserved quantity
\begin{equation}
\frac{\partial L}{\partial \dot{u}}=\frac{2}{\omega ^{2}}\dot{u}=\frac{3}{
\omega ^{2}}\sqrt{aV(\varphi )}(\dot{a}+a\dot{\varphi}).  \label{eq10}
\end{equation}

We can now easily obtain a general exact solution
\begin{equation}
u=u_{1}t+u_{2}\quad ,\quad v=v_{1}\sin (\omega t+v_{2}).  \label{eq11}
\end{equation}

Thus, we have four integration constants ($u_{1\text{, }}u_{2}$,
$v_{1}$, $ v_{2})$ and two free parameters ($\omega $ and
$\varphi_{\infty }$, which are in turn derived from $A$ and $B$ in the
potential).

First of all, let us set $u_{2}=v_{2}=0$. This is more than needed to fix a
time origin, i.e., $a(0)=0$, but it is important to select the tracking
solution. At the end of Sec. III we shall consider a weaker condition.

The second step consists in fixing the time scale; we pose, then, $t_{0}=1$.

Finally, let us denote by ${\cal H}_0$ the present value of the Hubble
parameter [but measured in units of (age of the universe)$^{-1}$],
which in our units is of order 1, even if it does not coincide with
the value of the usual $h$.

When all these conditions are imposed we obtain the following expressions
\begin{eqnarray}
a^{3} &=&\frac{(3{\cal H}_{0}-2)\csc ^{2}\omega \sin ^{2}\omega
t+(2\omega
\cot
\omega -3{\cal H}_{0})t^{2}}{2(\omega \cot \omega -1)},  \label{eq12} \\
\varphi  &=&\sqrt{\frac{2}{3}}\log \left\{ \left( t\sin \omega \sqrt{
(3{\cal H}_{0}-2\omega \cot \omega )(\omega \cot \omega -1)}\right.
\right.
\nonumber \\
&&\left. -\sqrt{(2-3{\cal H}_{0})(1-\omega \cot \omega )}\sin \omega
t\right)
\left/ t\sin \omega \sqrt{(3{\cal H}_{0}-2\omega \cot \omega )(\omega \cot \omega
-1)}\right.   \nonumber \\
&&\left. +\sqrt{(2-3{\cal H}_{0})(1-\omega \cot \omega )}\sin \omega
t\right) +\varphi _{\infty },  \label{eq13} \\ H &=&\frac{2t(3{\cal
H}_{0}-2\omega \cot \omega )-(3{\cal H}_{0}-2)\omega
\csc
^{2}\omega
\sin 2\omega t}{3t^{2}(3{\cal H}_{0}-2\omega \cot \omega )-3(3{\cal H}_{0}-2)\csc
^{2}\omega \sin ^{2}\omega t},  \label{eq14} \\
V(t) &=&\frac{4(3{\cal H}_{0}-2)\omega \csc ^{2}\omega \sin
^{2}\omega t}{ 3t^{2}(2\omega \cot \omega -3{\cal H}_{0})+3(3{\cal
H}_{0}-2)\csc
^{2}\omega
\sin
^{2}\omega t},  \label{eq15} \\
\dot{\varphi}^{2} &=&\frac{8(3{\cal H}_{0}-2)\sin \omega (3{\cal H}_{0}\sin \omega
-2\cos
\omega )(\sin \omega t-\omega t\cos \omega t)^{2}}{3(t^{2}(\omega \sin
2\omega -3{\cal H}_{0}\sin ^{2}\omega )+(3{\cal H}_{0}-2)\sin
^{2}\omega t)^{2}}.
\label{eq16}
\end{eqnarray}

From Eq. (\ref{eq12}) it is clear that $a$, $\dot{a}$, and $\ddot{a}$
are all oscillating functions. Actually, $\ddot{a}$ still oscillates
to infinity, when it behaves like
\begin{equation}
\ddot{a} \propto  \frac{2^{\frac{2}{3}}\,\left( -2+3\,{\cal H}_{0}\right) \,{\omega }
^{2}\,{\csc (\omega )}^{2}}{3\,{\left( -1+\omega \,\cot (\omega )\right) }^{
\frac{1}{3}}\,{\left( t^{2}\,\left( -3\,{\cal H}_{0}+2\,\omega \,\cot (\omega
)\right) \right) }^{\frac{2}{3}}}\cos (2\,\omega \,t)\,.
\label{eq12a}
\end{equation}
It is interesting to note that the right-hand side in Eq.
(\ref{eq12a}) is a damped oscillating function, unless ${\cal H}_{0}
={\frac{2}{3}}$ ; however, since ${\cal H}_{0} $ is not given by the
usual $h$ parameter here, as noted before, such a value is forbidden
by the recent observational data.

From the equations above it is straightforward to derive the
expressions for $\Omega _{\varphi }\equiv \rho _{\varphi }/\left(
3H^{2}\right) $ and $ w_{\varphi }\equiv p_{\varphi }/\rho _{\varphi
}$ ($p_{\varphi }$ being the pressure connected to the scalar field),
which are not given explicitly here, since they are really messy. It
is also clear that $\varphi _{\infty }$ only enters into $\varphi
\left( t\right) $, and determines its asymptotic value. We are, thus,
left with $\omega $ and ${\cal H}_{0}$ as free parameters.

\section{Analysis of the solution}

First of all, let us observe that Eq. (\ref{eq14}), no matter which are the
chosen values, shows that $H(t\rightarrow \infty )\approx 2/\left( 3t\right)
$. That is, the behavior in the future approaches pure dust, and the
acceleration, if ever present, is switched off. This, of course, does not
mean very much, until we do prove that the model can match the observations.

As we are not interested in best fits with real observational data,
but rather in comparing the behavior of our model with a pure $\Lambda
$-term model, it is sufficient to choose reasonable values for $\omega
$ and ${\cal H}_{0}$. Since ${\cal H}_{0}\approx 1$, let us set this
value for the moment and look at the plot of $\Omega _{0\varphi }$
versus $\omega $ (see Fig. 1). We discover that large $\omega $ values
yield nonphysical results. The best choice seems to be $\omega <2$.
Provided this limitation is set, the value of $\Omega
_{0\varphi }$ is actually not very sensible to $\omega ,$ so that this
parameter is in fact almost degenerate.

Finally, we choose (in our units)
\begin{equation}
{\cal H}_{0}=0.95,\quad \omega =0.3\qquad \Longrightarrow \qquad
\Omega
_{0\varphi }\approx 0.72,  \label{eq17}
\end{equation}
which seems quite reasonable. Let us also consider a cosmological
constant model with
\begin{equation}
{\cal H}_{0}=1,\quad \Omega _{\Lambda }\equiv \Omega _{0\varphi
}\approx 0.76.
\label{eq18}
\end{equation}
We have to insist on the fact that these values are just indicative. What we
want to say is that, after considering one of the two models as fiducial,
they are perfectly equivalent if one takes rather small differences in the
free parameters. Indeed, comparing the two luminosity distance curves we
obtain the strict overlapping in Fig. 2.

It should be clear from this that the two models will stay
undistinguishable not only for the present observations with SNIa, but
also with near future improvements, like those supposed to come from
Supernova/Acceleration Probe (SNAP) mission \cite{15}. Only a
tremendous advance in the precision could make the difference.

A comparison of the two models with real data on peculiar velocities
of the galaxies is also made in Ref. \cite{16}, and we again find good
agreement and impossibility to distinguish.

A study of the consequences on the cosmic microwave background
radiation (CMBR) spectrum is now in preparation, but a first check
seems to give the same type of answer \cite{17}.

In order to complete the discussion, let us look at the plot of the
acceleration $\ddot{a}(t)$ (Fig. 3). We see that it evolves towards
alternative periods of accelerations and decelerations, which justifies the
title of this paper. With our choice for the parameters, all this happens in
the far future, and we now live in the first period of acceleration. The
reason why it is so is connected to the coincidence problem. In any case, we
have a hint on a possible way to face the problem. It may be that a more
sophisticated model gives a less steep damping of the oscillations.

Let us now show the plots of $\log _{10}\rho _{\varphi }$ versus $\log
_{10}a $ and $w\equiv w_{\varphi }$ versus $\log _{10}a$ (see Figs. 4
and 5). After the above discussion, the first plot should be
self-explicative: it has the shape usually obtained with qualitative
arguments, but in this case it is obtained by exact integration, which
seems noteworthy to us. In the second plot the present time is marked
by the zero point and seems to be very special. We can imagine to be
in the far past or in the far future, and make the same analysis with
largely different values for $\Omega_{0\varphi }$. Considering such
values in the range $\{0.01\div 0.93\}$, we obtain the overlap in Fig.
6, showing that the coincidence is only apparent.

Last but not least, we can show that the examined solution also exhibits the
socalled tracking behavior.

So, let us release the condition $u_{2}=v_{2}=0$, and substitute it
with the more general $u_{2}=v_{2}=x$, which is the minimum necessary
to have $a(0)=0$. Then, let us also consider all the other constraints
introduced in Sec. II. Allowing $x$ to vary over a very narrow range
around zero, Fig. 4 is modified into Fig. 7, showing that a large set
of initial conditions eventually evolves towards the same final
behavior. Even if $x$ is increased, the qualitative behavior does not
change much, but the time needed for reaching the tracking solution is
now larger, leaving still opened a sort of fine tuning problem. In any
case, let us observe that $x$ is only a mathematical object, without
any obvious physical interpretation. The relevant physical quantity
$\rho_\varphi$ is not fine tuned, as we have shown.

\section{Conclusions}

The model presented here, in our opinion, shows a possible solution for all
the main problems posed by the dark energy theory, at least from a
mathematical point of view.

The physical attitude should be however more careful. Though very simple and
reasonable, the potential proposed here has at the moment not satisfactory
justification, independent of this model.

We have seen that, were this the real situation, it could be probably
impossible to discriminate from a $\Lambda $-term model with any
realistic plan of observations.\footnote{ Note that this tendency also
results from the very recent WMAP data \cite{18}
\par
{}}. As for the switch of the acceleration in the future, it is of
course unobservable in principle!

The main result of this paper seems to be an indication of a possible
way to solve the problem of switching off the acceleration by means of
minimal tools. It also suggests that oscillations of the scalar field
could be an interesting field of study also in a context very far from
inflation and reheating. (For some considerations about this, for
example, see Refs. \cite{19}, \cite{20}.)

A last remark has to be addressed to the fact that the radiation
dominated era is not recovered at early times. As a matter of fact,
this is due to the model itself and the way it is built. Adding a
radiation component is not simple on an analytical ground, since in
fact it is not possible to find a Noether symmetry and, therefore, an
analytical exact general solution. This means that including
radiation, so extending the model backwards, at least needs numerical
techniques, if one also wants to retain the other two components, dust
and scalar field.

\newpage

\begin{figure}[tbp]
\includegraphics{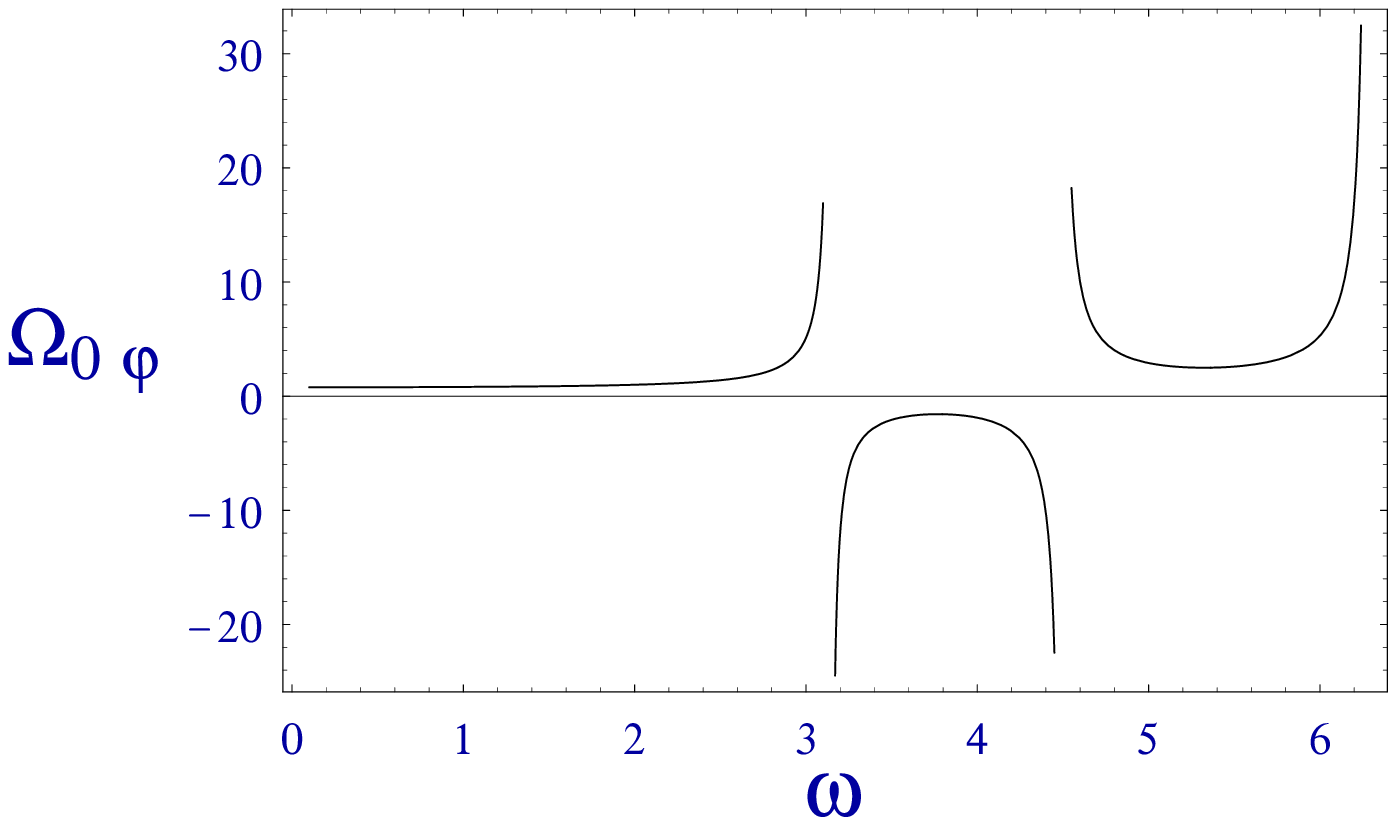}
\caption{The value of $\omega$ must be $\ll 1$ in order to obtain physical
values for the scalar field density.}
\label{Fig1}
\end{figure}

\begin{figure}[tbp]
\includegraphics{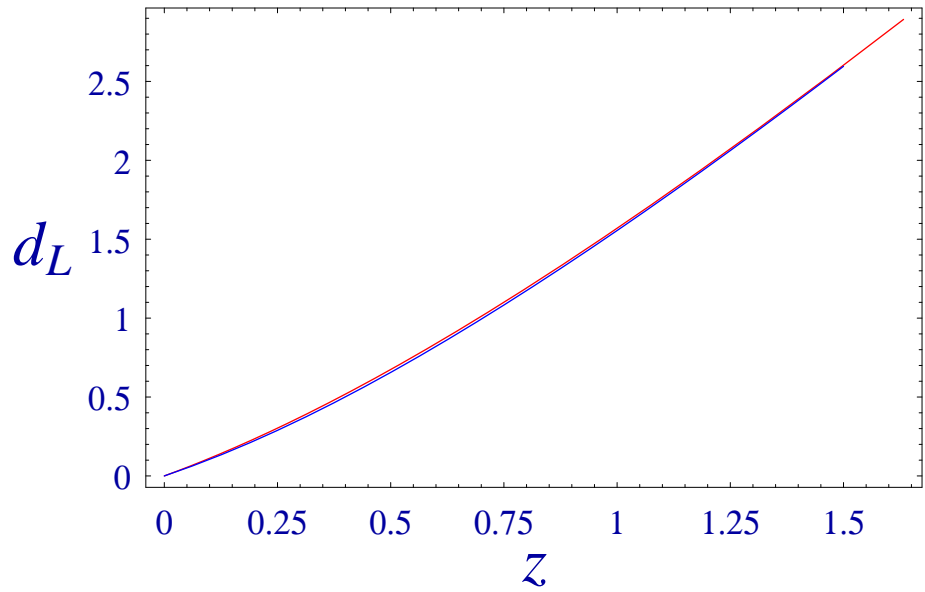}
\caption{Comparison of the luminosity distance for a $\Lambda$-term model
(blue line), setting ${\cal H}_0= 1$ and $\Omega_{0\phi} = 0.76$ ,
with our model (red), setting ${\cal H}_0= 0.95$ and $\Omega_{0\phi}
= 0.7$. The two curves are practically undistinguishable. }
\label{Fig2}
\end{figure}

\begin{figure}[tbp]
\includegraphics{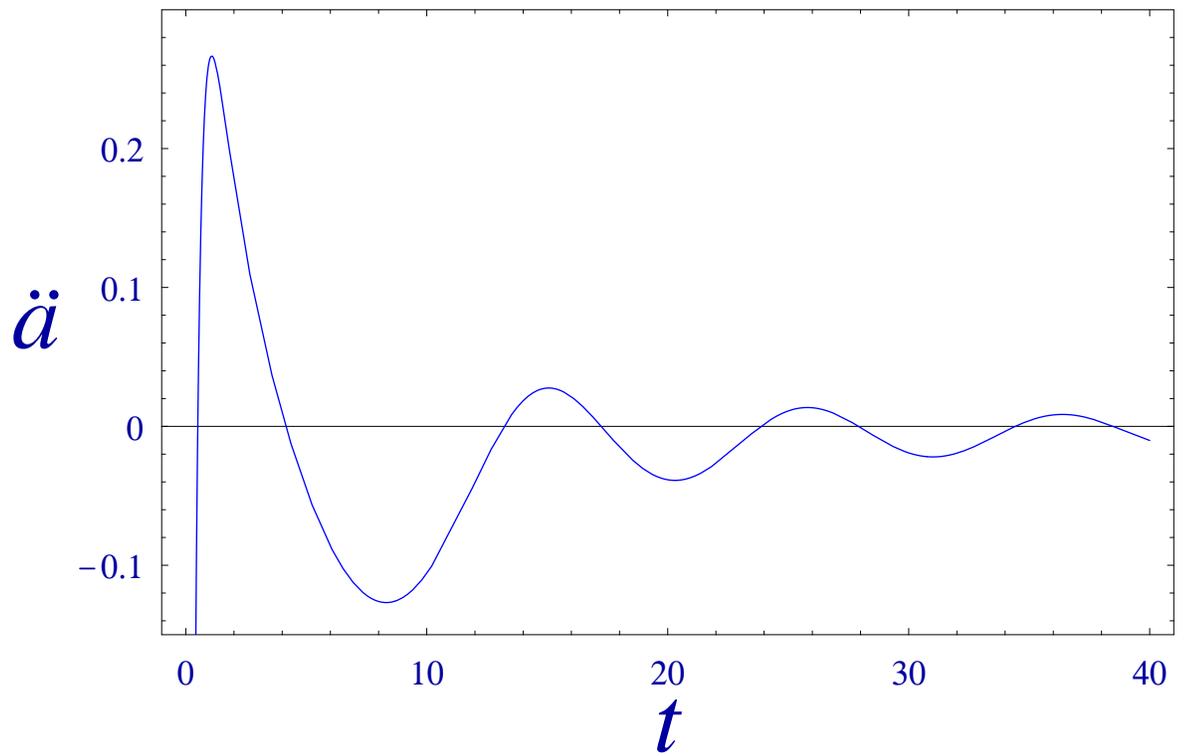}
\caption{ Evolution of the acceleration with time. The present time is $t_0 =
1$. In the future the acceleration shows damped oscillation, with zero
limit.}
\label{Fig3}
\end{figure}

\begin{figure}[tbp]
\includegraphics{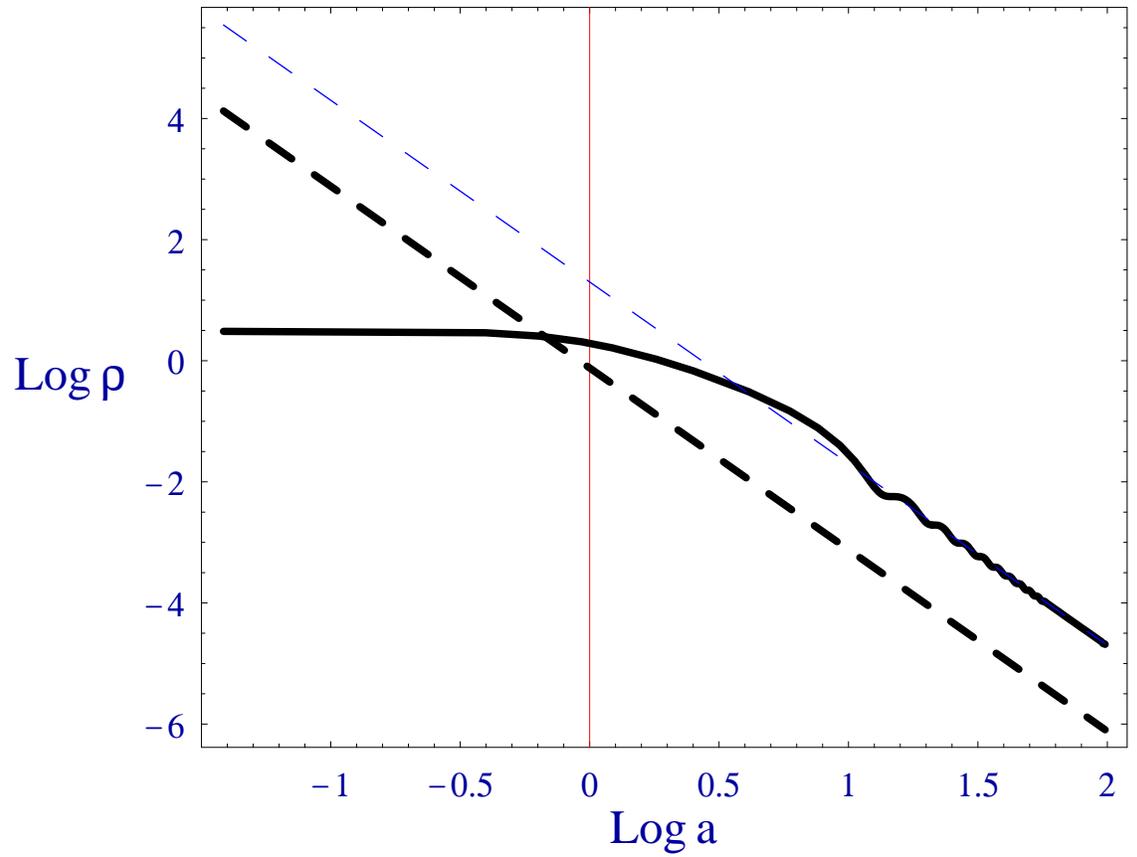}
\caption{ Scalar field density versus scale factor. The (black) thick dashed
line is the matter density. The (blue) thin dashed line is drawn for
reference and shows that the asymptotic value of the state equation
for the scalar field is the same as matter.}
\label{Fig4}
\end{figure}

\begin{figure}[tbp]
\includegraphics{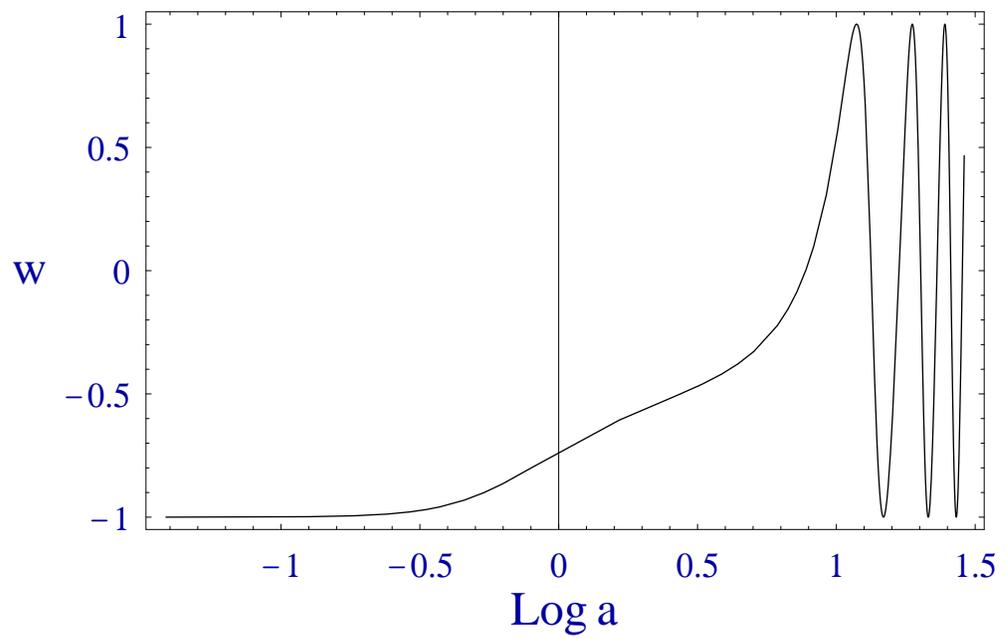}
\caption{ The state equation of the scalar field versus the scale factor.
Apparently, the present time (marked by the Y axis) is in the middle
of a phase transition between $w = -1$ and oscillations. The next
figure clarifies the situation.}
\label{Fig5}
\end{figure}

\begin{figure}[tbp]
\includegraphics{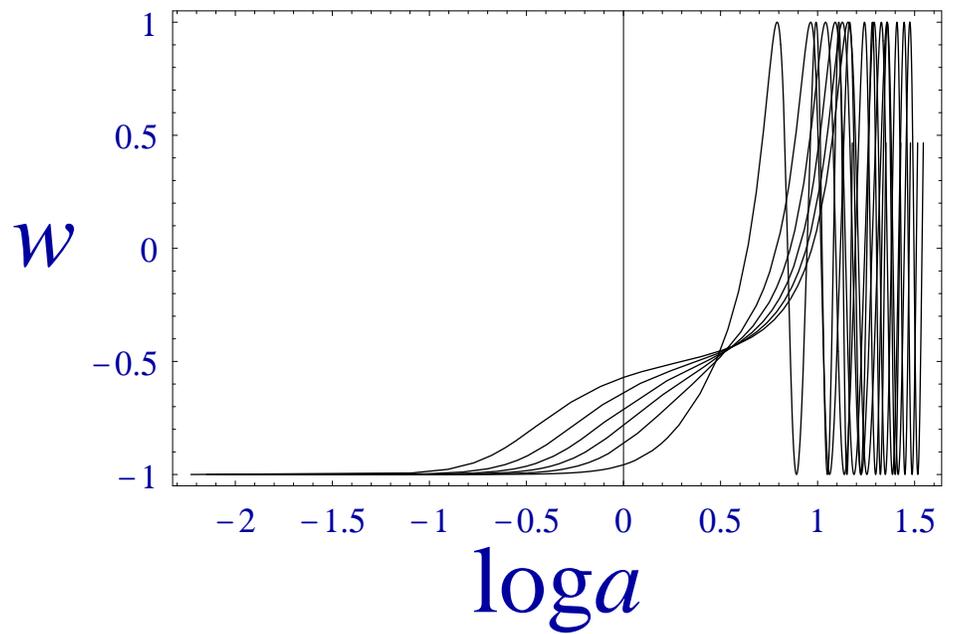}
\caption{ Same as Fig. 5, but now several ``present'' values of $\Omega_{0\phi}$
are used, in the range $0.01 \div 0.93$. This shows that the plots are
very similar and that the coincidence is only apparent.}
\label{Fig6}
\end{figure}

\begin{figure}[tbp]
\includegraphics{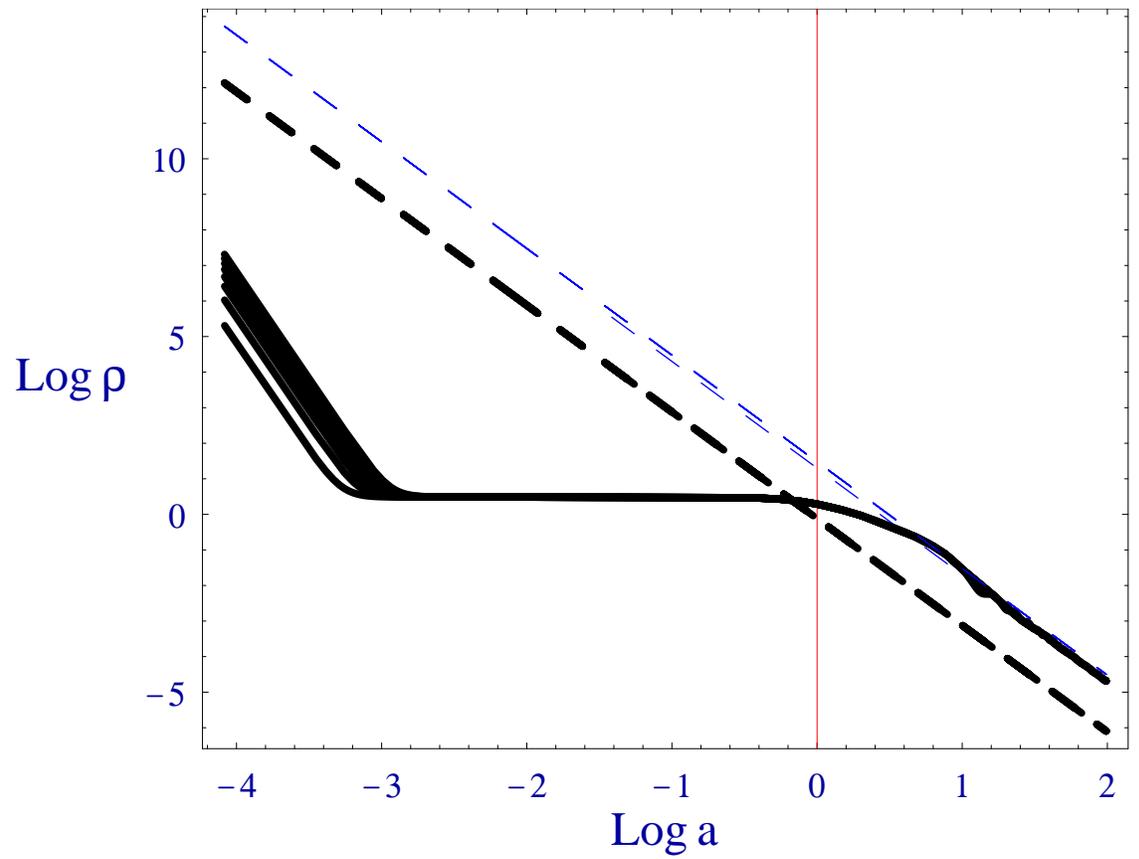}
\caption{Same as Fig. 4, but with the tracking behavior, due to the
variation of the $x$ parameter (see text).}
\label{Fig7}
\end{figure}

\end{document}